\documentstyle[twoside,fleqn,espcrc2]{article}

\newcommand{\AmS}{{\protect\the\textfont2
  A\kern-.1667em\lower.5ex\hbox{M}\kern-.125emS}}
\title{Search for Effective Lattice Action of Pure QCD}

\author{QCDTARO Collaboration\\
	Ph.~de~Forcrand \address{Swiss Center for Scientific Computing (SCSC), 
	  ETH-Z\"urich, CH-8092 Z\"urich , Switzerland},
	M.~Fujisaki \address{High Performance Computing Group, Fujitsu Ltd.,
	  Mihama-ku, Chiba 261, Japan},
	T.~Hashimoto \address{Department of Applied Physics, Fukui University,
	  Fukui 910, Japan},
	S.~Hioki \address{Department of Physics, Hiroshima University,
	  Higashi-Hiroshima 739, Japan},
	H.~Matsufuru $^{\rm d}$,
	O.~Miyamura $^{\rm d}$,
	A.~Nakamura \address{Faculty of Education, Yamagata University,
	  Yamagata 990, Japan},
	M.~Okuda ${\rm ^b}$,
	I.~-O.~Stamatescu \address{FESt Heidelberg and Institut f\"ur 
	  Theoretische Physik, Universit\"at Heidelberg, D-69120 Heidelberg, 
	  Germany},
	T.~Takaishi ${\rm ^a}$ and 
	Y.~Tago \address{Computational Science Research Laboratory, Fujitsu Ltd.,
	  Mihama-ku, Chiba261, Japan}
}       
\begin{document}

\begin{abstract}
We study a coupling flow of pure QCD gauge system by using the
Monte Carlo Renormalization Group method.
A rough location of the renormalized trajectory in two coupling space
is obtained.
Also we compare 4 different actions; (a)standard Wilson,
(b)Symanzik's, (c)Iwasaki's and (d)QCDTARO's.
The rotational symmetry is restored better as an action gets close to
the renormalized trajectory.
\end{abstract}

\maketitle
\thispagestyle{empty}

\section{Introduction}

In order to achieve the continuum physics in lattice gauge theory,
a lattice spacing $a$ should be small enough.
However the scaling violation of Wilson action at presently
accessible $a$ is clearly seen \cite{Ukawa93,QCDTARO93,Lepage96}.
Since the cost of a Monte Carlo Simulation of QCD can estimated
as $\propto a^{-6}$ \cite{Lepage96}, it seems difficult to 
adopt smaller $a$ without a surprising computer power-up or 
improvement of the algorithm. 

Recently "improved" actions have been extensively applied to resolve
the scaling problems \cite{This}.
Although in these studies, several types of actions are proposed and used. 
What is the most optimal choice or what is a "perfect" action is still
an open question.

In this paper, we analyze the coupling flow of pure SU(3) gauge theory
by the Monte Carlo Renormalization Group (MCRG) method.
And we compare several actions, which are widely used in actual simulations,
by investigating the rotational symmetry.

\section{Improved Action}

In this paper we consider 2-coupling space in which the 
$1\times 1$ simple plaquette and $1\times 2$ rectangle loop 
are included. 
The action $S$ is,
\begin{eqnarray}
S & = & \beta_{11} \sum_{plaq} {\rm Re Tr} (1 - {1 \over 3} U_{plaq}) \\
  & + & \beta_{12} \sum_{rect} {\rm Re Tr} (1 - {1 \over 3} U_{rect}),\nonumber
\label{eq:action}
\end{eqnarray}
where
\vspace{-5mm}
$$
U_{plaq} = 
\begin{picture}(20,20)
{\thicklines
\put(0,0){\vector(1,0){20}}
\put(20,0){\vector(0,1){20}}
\put(20,20){\vector(-1,0){20}}
\put(0,20){\vector(0,-1){20}}
}
\end{picture}
\hspace{5mm},\hspace{5mm}
U_{rect} = 
\begin{picture}(40,20)
{\thicklines
\put(0,0){\vector(1,0){20}}
\put(20,0){\vector(1,0){20}}
\put(20,0){\circle*{2}}
\put(40,0){\vector(0,1){20}}
\put(40,20){\vector(-1,0){20}}
\put(20,20){\vector(-1,0){20}}
\put(20,20){\circle*{2}}
\put(0,20){\vector(0,-1){20}}
}.
\end{picture} 
$$

Although there are several arguments about the necessity of other loops
as possible candidates of an action \cite{Lepage96,Takaishi96},
actions with two couplings are widely used as improved ones 
in actual lattice QCD simulations \cite{This}.
Here, we restrict ourselves on a 2 coupling space and
our purpose of this paper is to propose an optimal choice of action in
2 coupling space.
Inclusion of other couplings is now in progress and will be presented 
elsewhere.

For actions with 2 couplings, we consider 4 candidates which are widely used;
(a) Wilson \cite{Wilson}, (b) Symanzik \cite{Symanzik},
(c) Iwasaki \cite{Iwasaki}, (d) QCDTARO \cite{Takaishi96}.

$\beta$s (or the ratio $\gamma_{12}\equiv\beta_{12} / \beta_{11}$) 
in these actions are listed in Table~\ref{tab:betas}.
\vspace{-5mm}
\begin{table}[hbt]
\caption{couplings in each action}
\label{tab:betas}
\begin{tabular}{lrrr}
\hline
 & $\gamma_{12}$\ \  & $\beta_{11}$\ \ \ \ \ \  & $\beta_{12}$\ \ \ \ \ \  \\
\hline
(a)Wilson           &  0            & & \\
(b)Symanzik         &  -0.05        & & \\
(c)Iwasaki           &  -0.091       & & \\
(d)QCDTARO          &               & 6.1564(53) & -0.6241(23) \\
             &       & 7.986(13)  & -0.9169(41) \\
\hline
\end{tabular}
\end{table}
\vspace{-1.2cm}

\section{Coupling Flow by MCRG}

First we follow the coupling flow by MCRG. 
As MCRG method, here we adopt Swendsen's factor 2 blocking scheme 
\cite{Swendsen}.
For details of this blocking, see \cite{QCDTARO93,Takaishi96}. 

\subsection{Schwinger-Dyson method for determination of effective action}

From the original configurations $\{U\}$ which is generated by an action $S$
with $(\beta_{11},\beta_{12})$ in eq.(\ref{eq:action})
we obtain the blocked ones
$\{U'\}$ after the above blocking procedure.
When the blocked configurations $\{U'\}$ can be considered as generated
configurations by an action $S'$ with $(\beta'_{11},\beta'_{12})$ 
in eq.(\ref{eq:action}),
an action $S'$ is an effective action of the blocked configurations, 
and 
\begin{equation}
(\beta_{11},\beta_{12}) \rightarrow (\beta'_{11},\beta'_{12})
\label{eq:flow}
\end{equation}
can be regarded as the coupling flow associated with this blocking.

For the determination of an effective action, here we use a 
Schwinger-Dyson method \cite{Okawa}.

This method is simply based on the following identity. 
For a link $U_l$, we consider a quantity;

\begin{equation}
F =  \prod_{\bar l} \int dU_{\bar l} {\rm Im}{\rm Tr}(\lambda^a U_l G_l^{\alpha})
e^{2 {\rm ReTr}(U_l G_l)},
\end{equation}
\noindent
where $\bar l$ means link except for $l$. 
$G_l$ is a sum of staples $G_l^{\alpha}$ for link $l$ as 
$G_l = \sum_{\alpha} {\beta \over 6} G_l^{\alpha}$.
For infinitesimal transformation;
$U_l \rightarrow (1+i\epsilon\lambda^{\alpha})U_l, $ 
invariance of $F$ leads to 

\vspace*{-5mm}
\begin{eqnarray}
{dF \over d\epsilon} &=& \prod_{\bar l} \int dU_{\bar l}
[ {\rm Re} {\rm Tr}((\lambda^a)^2 U_l G_l^{\alpha}) \\
&+& {\rm Im} {\rm Tr}(\lambda^a U_l G_l^{\alpha}) {\rm Im} {\rm Tr}(\lambda^a U_l G_l)] \nonumber \\
&\times& e^{2 {\rm ReTr}(U_l G_l)} = 0 \ \ \nonumber. 
\end{eqnarray}

Using a well-known formula of $\lambda$-matrix and summation
 over $a$ and integral over link $U_l$ , we have

\vspace*{-5mm}
\begin{eqnarray}
& & {8 \over 3}{\rm Re}<{\rm Tr}(U_l G_l^{\alpha})> =  \\ 
& & \sum_{\gamma} {\beta \over 6}\{
-  {\rm Re} <{\rm Tr}(U_lG_l^{\alpha}U_lG_l^{\gamma})>  \nonumber\\
& & +  {\rm Re} <{\rm Tr}(G_l^{\alpha}(G_l^{\gamma})^{\dagger})> \nonumber\\
& & - {1 \over 3}{\rm Re} <{\rm Im Tr}(U_lG_l^{\alpha})
{\rm Im Tr}(U_lG_l^{\gamma})>\} \ \ \nonumber. 
\end{eqnarray}

We then apply this equation to the blocked configurations. 

\subsection{Lattice Simulation and Results}

We use $8^4$ lattice size and about 2000 configurations separated by
every 10 sweeps are used.
For the starting point $(\beta_{11},\beta_{12})$, 
we try (a)-(d) coupling sets in 
Table~\ref{tab:betas} and other points are also tested to search the
renormalized trajectory(RT) in detail.
All simulations have been done on CRAY J90 at Information Processing Center,
Hiroshima University and on VPP500 at KEK (National Laboratory for High Energy
Physics). 

In Fig.1, the result of coupling flows in $(\beta_{11},\beta_{12})$
plane is shown. 
Arrow denotes the measured coupling flow in eq.(\ref{eq:flow}).

Ending points seem lie on the universal line which can be thought as the
RT of this blocking scheme.
For comparison, we prepare 2 sets of (a)-(d) points at almost the same scale;
$a^2\sigma \simeq 1.51(6)$ and $a^2\sigma \simeq 0.82(6)$.
Among (a)-(d) at almost the same scale, 
(d) is the closest to this RT and (c) is the next.
 
\section{Rotational Invariance}

In order to see the improvement of the action, here we check the rotational
invariance by measuring the heavy quark potential.
In this paper, we compare 4 actions in Table~\ref{tab:betas} at 
$a^2\sigma \simeq 1.51(6)$.

In Fig.\ref{fig:potential}, the result is shown.
Since the lattice spacing at this scale is rather large,
rotational symmetry can not be expected to be restored on the standard
Wilson action, which is clearly seen in Fig.\ref{fig:potential}(a).
For Symanzik action in Fig.\ref{fig:potential}(b), 
the situation is similar to Wilson case.
On the other hand for Iwasaki and QCDTARO cases in 
Figs.\ref{fig:potential}(c) and (d), the rotational invariance can be seen.
It is also noted that errors become small as the action gets close to RT.

In order to compare quantitatively, we fit the measured potential 
$V(R_i)$ ($R_i$ is the measured point) with
the linear plus coulomb term as $ f(R) = a + b R + c/R $ and evaluate
the effectiveness of this fit by
\begin{equation}
\label{eq:chi}
 \chi^2 \equiv \sum_i ( V(R_i) - f(R_i) )^2
\end{equation}

\vspace{-10mm}
\begin{table}[hbt]
\label{tab:ci}
\begin{tabular}{ccccc}
\hline
 & Wilson & Symanzik & Iwasaki & QCDTARO \\
\hline
$\chi$ & .302   & .228 & .092 & .076 \\
\hline
\end{tabular}
\end{table}
\vspace{-.5cm}

From these results, rotational symmetry can be restored better as the action
gets close to RT.

\section*{AKNOWLEDGEMENTS}

To attend this conference,
S.H is supported by the Grant-in-Aid for Scientific Research 
of the Ministry of Education No.08640379.
 
\setcounter{figure}{1}
\begin{figure}[htb]
\unitlength 1.0mm
\begin{picture}(60,52)(-10,-10)
\def\xw{60.000000} \def\yw{40.000000}
\put(50,3){\large $R/a$}
\put(3,35){\large (a)}
\put(-13,33){\large $aV(R)$}
\def\errorbar2#1#2#3#4#5#6{
\put(#1,#2){\line(0,1){#3}}
\put(#1,#2){\line(0,-1){#3}}
\put(#4,#5){\line(1,0){1}}
\put(#4,#6){\line(1,0){1}} }
\def\err3#1#2#3#4#5#6{
\put(#1,#2){\vector(0,1){#5}}
\put(#1,#2){\line(0,-1){#3}}
\put(#4,#6){\line(1,0){1}} }
\errorbar2{15.000000}{18.551800}{0.077501}{14.500000}{18.629300}{18.474299}
\errorbar2{21.210001}{24.134901}{0.257437}{20.710001}{24.392338}{23.877464}
\errorbar2{25.980000}{28.108700}{3.227480}{25.480000}{31.336180}{24.881220}
\errorbar2{30.000000}{26.532499}{1.102310}{29.500000}{27.634809}{25.430189}
\errorbar2{33.540001}{31.253601}{3.783910}{33.040001}{35.037511}{27.469691}
\err3{42.420002}{34.607201}{28.910601}{41.920002}{4.4}{5.696599}
\err3{45.000000}{33.790100}{19.161900}{44.500000}{5.2}{14.628200}
\put(15.000000,18.551800){\circle{1.000000}}
\put(21.210001,24.134901){\circle{1.000000}}
\put(25.980000,28.108700){\circle{1.000000}}
\put(30.000000,26.532499){\circle{1.000000}}
\put(33.540001,31.253601){\circle{1.000000}}
\put(42.420002,34.607201){\circle{1.000000}}
\put(45.000000,33.790100){\circle{1.000000}}
\put(7.500000,3.224539){\circle*{0.100000}}
\put(9.000000,8.139149){\circle*{0.100000}}
\put(10.500001,11.745385){\circle*{0.100000}}
\put(12.000001,14.533890){\circle*{0.100000}}
\put(13.500002,16.777241){\circle*{0.100000}}
\put(15.000002,18.638981){\circle*{0.100000}}
\put(16.500002,20.223190){\circle*{0.100000}}
\put(18.000002,21.599247){\circle*{0.100000}}
\put(19.500004,22.815191){\circle*{0.100000}}
\put(21.000004,23.905329){\circle*{0.100000}}
\put(22.500004,24.894821){\circle*{0.100000}}
\put(24.000004,25.802542){\circle*{0.100000}}
\put(25.500004,26.642920){\circle*{0.100000}}
\put(27.000004,27.427179){\circle*{0.100000}}
\put(28.500006,28.164181){\circle*{0.100000}}
\put(30.000004,28.861012){\circle*{0.100000}}
\put(31.500002,29.523411){\circle*{0.100000}}
\put(33.000000,30.156076){\circle*{0.100000}}
\put(34.500000,30.762882){\circle*{0.100000}}
\put(35.999996,31.347065){\circle*{0.100000}}
\put(37.499996,31.911337){\circle*{0.100000}}
\put(38.999996,32.457996){\circle*{0.100000}}
\put(40.499992,32.989002){\circle*{0.100000}}
\put(41.999992,33.506027){\circle*{0.100000}}
\put(43.499992,34.010521){\circle*{0.100000}}
\put(44.999989,34.503738){\circle*{0.100000}}
\put(46.499989,34.986763){\circle*{0.100000}}
\put(47.999985,35.460556){\circle*{0.100000}}
\put(49.499985,35.925957){\circle*{0.100000}}
\put(50.999985,36.383709){\circle*{0.100000}}
\put(52.499981,36.834457){\circle*{0.100000}}
\put(0.000000,0){\line(0,1){1}}
\put(15.000000,0){\line(0,1){1}}
\put(30.000000,0){\line(0,1){1}}
\put(45.000000,0){\line(0,1){1}}
\put(0.000000,0){\line(0,1){1}}
\put(0.000000,-3.5){\normalsize 0}
\put(15.000000,-3.5){\normalsize 1}
\put(30.000000,-3.5){\normalsize 2}
\put(45.000000,-3.5){\normalsize 3}
\put(0,10.000000){\line(1,0){1}}
\put(\xw,10.000000){\line(-1,0){1}}
\put(0,20.000000){\line(1,0){1}}
\put(\xw,20.000000){\line(-1,0){1}}
\put(0,30.000000){\line(1,0){1}}
\put(\xw,30.000000){\line(-1,0){1}}
\put(0,40.000000){\line(1,0){1}}
\put(\xw,40.000000){\line(-1,0){1}}
\put(0,10.000000){\line(1,0){1}}
\put(\xw,10.000000){\line(-1,0){1}}
\put(-1.900000,8.500000){\normalsize 0}
\put(-1.900000,18.500000){\normalsize 1}
\put(-1.900000,28.500000){\normalsize 2}
\put(-1.900000,38.500000){\normalsize 3}
{\linethickness{0.25mm}
\put(  0,  0){\line(1,0){\xw}}
\put(  0,\yw){\line(1,0){\xw}}
\put(\xw,  0){\line(0,1){\yw}}
\put(  0,  0){\line(0,1){\yw}} }
\end{picture} 
\begin{picture}(60,45)(-10,-10)
\def\xw{60.000000} \def\yw{40.000000}
\put(50,3){\large $R/a$}
\put(3,35){\large (b)}
\put(-13,33){\large $aV(R)$}
\def\errorbar2#1#2#3#4#5#6{
\put(#1,#2){\line(0,1){#3}}
\put(#1,#2){\line(0,-1){#3}}
\put(#4,#5){\line(1,0){1}}
\put(#4,#6){\line(1,0){1}} }
\errorbar2{15.000000}{17.939779}{0.068227}{14.500000}{18.008006}{17.871552}
\errorbar2{21.210001}{22.458399}{0.171286}{20.710001}{22.629685}{22.287113}
\errorbar2{25.980000}{25.834900}{0.834202}{25.480000}{26.669102}{25.000698}
\errorbar2{30.000000}{25.738100}{0.455413}{29.500000}{26.193513}{25.282687}
\errorbar2{33.540001}{28.376499}{1.714080}{33.040001}{30.090579}{26.662419}
\errorbar2{42.420002}{33.458500}{20.337200}{41.920002}{53.795700}{13.121300}
\errorbar2{45.000000}{31.960499}{10.614901}{44.500000}{42.575399}{21.345598}
\put(15.000000,17.939779){\circle{1.000000}}
\put(21.210001,22.458399){\circle{1.000000}}
\put(25.980000,25.834900){\circle{1.000000}}
\put(30.000000,25.738100){\circle{1.000000}}
\put(33.540001,28.376499){\circle{1.000000}}
\put(42.420002,33.458500){\circle{1.000000}}
\put(45.000000,31.960499){\circle{1.000000}}
\put(7.500000,7.111330){\circle*{0.100000}}
\put(9.000000,10.431926){\circle*{0.100000}}
\put(10.500001,12.935975){\circle*{0.100000}}
\put(12.000001,14.929683){\circle*{0.100000}}
\put(13.500002,16.583162){\circle*{0.100000}}
\put(15.000002,17.998482){\circle*{0.100000}}
\put(16.500002,19.240595){\circle*{0.100000}}
\put(18.000002,20.352802){\circle*{0.100000}}
\put(19.500004,21.365084){\circle*{0.100000}}
\put(21.000004,22.298849){\circle*{0.100000}}
\put(22.500004,23.169806){\circle*{0.100000}}
\put(24.000004,23.989727){\circle*{0.100000}}
\put(25.500004,24.767620){\circle*{0.100000}}
\put(27.000004,25.510490){\circle*{0.100000}}
\put(28.500006,26.223866){\circle*{0.100000}}
\put(30.000004,26.912172){\circle*{0.100000}}
\put(31.500002,27.578991){\circle*{0.100000}}
\put(33.000000,28.227251){\circle*{0.100000}}
\put(34.500000,28.859373){\circle*{0.100000}}
\put(35.999996,29.477375){\circle*{0.100000}}
\put(37.499996,30.082952){\circle*{0.100000}}
\put(38.999996,30.677538){\circle*{0.100000}}
\put(40.499992,31.262352){\circle*{0.100000}}
\put(41.999992,31.838444){\circle*{0.100000}}
\put(43.499992,32.406715){\circle*{0.100000}}
\put(44.999989,32.967945){\circle*{0.100000}}
\put(46.499989,33.522816){\circle*{0.100000}}
\put(47.999985,34.071926){\circle*{0.100000}}
\put(49.499985,34.615799){\circle*{0.100000}}
\put(50.999985,35.154892){\circle*{0.100000}}
\put(52.499981,35.689625){\circle*{0.100000}}
\put(0.000000,0){\line(0,1){1}}
\put(15.000000,0){\line(0,1){1}}
\put(30.000000,0){\line(0,1){1}}
\put(45.000000,0){\line(0,1){1}}
\put(0.000000,0){\line(0,1){1}}
\put(0.000000,-3.5){\normalsize 0}
\put(15.000000,-3.5){\normalsize 1}
\put(30.000000,-3.5){\normalsize 2}
\put(45.000000,-3.5){\normalsize 3}
\put(0,10.000000){\line(1,0){1}}
\put(\xw,10.000000){\line(-1,0){1}}
\put(0,20.000000){\line(1,0){1}}
\put(\xw,20.000000){\line(-1,0){1}}
\put(0,30.000000){\line(1,0){1}}
\put(\xw,30.000000){\line(-1,0){1}}
\put(0,40.000000){\line(1,0){1}}
\put(\xw,40.000000){\line(-1,0){1}}
\put(0,10.000000){\line(1,0){1}}
\put(\xw,10.000000){\line(-1,0){1}}
\put(-1.900000,8.500000){\normalsize 0}
\put(-1.900000,18.500000){\normalsize 1}
\put(-1.900000,28.500000){\normalsize 2}
\put(-1.900000,38.500000){\normalsize 3}
{\linethickness{0.25mm}
\put(  0,  0){\line(1,0){\xw}}
\put(  0,\yw){\line(1,0){\xw}}
\put(\xw,  0){\line(0,1){\yw}}
\put(  0,  0){\line(0,1){\yw}} }
\end{picture} 
\begin{picture}(60,45)(-10,-10)
\def\xw{60.000000} \def\yw{40.000000}
\put(50,3){\large $R/a$}
\put(3,35){\large (c)}
\put(-13,33){\large $aV(R)$}
\def\errorbar2#1#2#3#4#5#6{
\put(#1,#2){\line(0,1){#3}}
\put(#1,#2){\line(0,-1){#3}}
\put(#4,#5){\line(1,0){1}}
\put(#4,#6){\line(1,0){1}} }
\errorbar2{15.000000}{17.748270}{0.104205}{14.500000}{17.852475}{17.644065}
\errorbar2{21.210001}{21.268000}{0.429437}{20.710001}{21.697437}{20.838563}
\errorbar2{25.980000}{23.626801}{1.300000}{25.480000}{24.926800}{22.326801}
\errorbar2{30.000000}{26.016300}{0.427134}{29.500000}{26.443434}{25.589166}
\errorbar2{33.540001}{27.956999}{1.258630}{33.040001}{29.215629}{26.698369}
\errorbar2{42.420002}{33.554901}{15.192400}{41.920002}{48.747301}{18.362501}
\errorbar2{45.000000}{33.716198}{6.923520}{44.500000}{40.639718}{26.792678}
\put(15.000000,17.748270){\circle{1.000000}}
\put(21.210001,21.268000){\circle{1.000000}}
\put(25.980000,23.626801){\circle{1.000000}}
\put(30.000000,26.016300){\circle{1.000000}}
\put(33.540001,27.956999){\circle{1.000000}}
\put(42.420002,33.554901){\circle{1.000000}}
\put(45.000000,33.716198){\circle{1.000000}}
\put(7.500000,13.508125){\circle*{0.100000}}
\put(9.000000,14.370506){\circle*{0.100000}}
\put(10.500001,15.221622){\circle*{0.100000}}
\put(12.000001,16.065695){\circle*{0.100000}}
\put(13.500002,16.905075){\circle*{0.100000}}
\put(15.000002,17.741169){\circle*{0.100000}}
\put(16.500002,18.574875){\circle*{0.100000}}
\put(18.000002,19.406788){\circle*{0.100000}}
\put(19.500004,20.237322){\circle*{0.100000}}
\put(21.000004,21.066772){\circle*{0.100000}}
\put(22.500004,21.895355){\circle*{0.100000}}
\put(24.000004,22.723236){\circle*{0.100000}}
\put(25.500004,23.550535){\circle*{0.100000}}
\put(27.000004,24.377354){\circle*{0.100000}}
\put(28.500006,25.203764){\circle*{0.100000}}
\put(30.000004,26.029827){\circle*{0.100000}}
\put(31.500002,26.855593){\circle*{0.100000}}
\put(33.000000,27.681103){\circle*{0.100000}}
\put(34.500000,28.506392){\circle*{0.100000}}
\put(35.999996,29.331484){\circle*{0.100000}}
\put(37.499996,30.156404){\circle*{0.100000}}
\put(38.999996,30.981176){\circle*{0.100000}}
\put(40.499992,31.805811){\circle*{0.100000}}
\put(41.999992,32.630325){\circle*{0.100000}}
\put(43.499992,33.454735){\circle*{0.100000}}
\put(44.999989,34.279045){\circle*{0.100000}}
\put(46.499989,35.103268){\circle*{0.100000}}
\put(47.999985,35.927406){\circle*{0.100000}}
\put(49.499985,36.751476){\circle*{0.100000}}
\put(50.999985,37.575485){\circle*{0.100000}}
\put(52.499981,38.399426){\circle*{0.100000}}
\put(0.000000,0){\line(0,1){1}}
\put(15.000000,0){\line(0,1){1}}
\put(30.000000,0){\line(0,1){1}}
\put(45.000000,0){\line(0,1){1}}
\put(0.000000,0){\line(0,1){1}}
\put(0.000000,-3.5){\normalsize 0}
\put(15.000000,-3.5){\normalsize 1}
\put(30.000000,-3.5){\normalsize 2}
\put(45.000000,-3.5){\normalsize 3}
\put(0,10.000000){\line(1,0){1}}
\put(\xw,10.000000){\line(-1,0){1}}
\put(0,20.000000){\line(1,0){1}}
\put(\xw,20.000000){\line(-1,0){1}}
\put(0,30.000000){\line(1,0){1}}
\put(\xw,30.000000){\line(-1,0){1}}
\put(0,40.000000){\line(1,0){1}}
\put(\xw,40.000000){\line(-1,0){1}}
\put(0,10.000000){\line(1,0){1}}
\put(\xw,10.000000){\line(-1,0){1}}
\put(-1.900000,8.500000){\normalsize 0}
\put(-1.900000,18.500000){\normalsize 1}
\put(-1.900000,28.500000){\normalsize 2}
\put(-1.900000,38.500000){\normalsize 3}
{\linethickness{0.25mm}
\put(  0,  0){\line(1,0){\xw}}
\put(  0,\yw){\line(1,0){\xw}}
\put(\xw,  0){\line(0,1){\yw}}
\put(  0,  0){\line(0,1){\yw}} }
\end{picture} 
\begin{picture}(60,45)(-10,-10)
\def\xw{60.000000} \def\yw{40.000000}
\put(50,3){\large $R/a$}
\put(3,35){\large (d)}
\put(-13,33){\large $aV(R)$}
\def\errorbar2#1#2#3#4#5#6{
\put(#1,#2){\line(0,1){#3}}
\put(#1,#2){\line(0,-1){#3}}
\put(#4,#5){\line(1,0){1}}
\put(#4,#6){\line(1,0){1}} }
\errorbar2{15.000000}{17.257950}{0.075278}{14.500000}{17.333227}{17.182672}
\errorbar2{21.210001}{20.344400}{0.316576}{20.710001}{20.660976}{20.027824}
\errorbar2{25.980000}{22.538799}{0.886458}{25.480000}{23.425257}{21.652341}
\errorbar2{30.000000}{25.027401}{1.660230}{29.500000}{26.687631}{23.367171}
\errorbar2{33.540001}{25.999201}{0.838252}{33.040001}{26.837453}{25.160949}
\errorbar2{42.420002}{30.065701}{3.643700}{41.920002}{33.709400}{26.422001}
\errorbar2{45.000000}{30.388401}{3.929360}{44.500000}{34.317761}{26.459041}
\put(15.000000,17.257950){\circle{1.000000}}
\put(21.210001,20.344400){\circle{1.000000}}
\put(25.980000,22.538799){\circle{1.000000}}
\put(30.000000,25.027401){\circle{1.000000}}
\put(33.540001,25.999201){\circle{1.000000}}
\put(42.420002,30.065701){\circle{1.000000}}
\put(45.000000,30.388401){\circle{1.000000}}
\put(7.500000,10.672970){\circle*{0.100000}}
\put(9.000000,12.464976){\circle*{0.100000}}
\put(10.500001,13.903771){\circle*{0.100000}}
\put(12.000001,15.121809){\circle*{0.100000}}
\put(13.500002,16.192675){\circle*{0.100000}}
\put(15.000002,17.160521){\circle*{0.100000}}
\put(16.500002,18.053444){\circle*{0.100000}}
\put(18.000002,18.890173){\circle*{0.100000}}
\put(19.500004,19.683678){\circle*{0.100000}}
\put(21.000004,20.443218){\circle*{0.100000}}
\put(22.500004,21.175591){\circle*{0.100000}}
\put(24.000004,21.885887){\circle*{0.100000}}
\put(25.500004,22.578003){\circle*{0.100000}}
\put(27.000004,23.254969){\circle*{0.100000}}
\put(28.500006,23.919178){\circle*{0.100000}}
\put(30.000004,24.572540){\circle*{0.100000}}
\put(31.500002,25.216610){\circle*{0.100000}}
\put(33.000000,25.852650){\circle*{0.100000}}
\put(34.500000,26.481710){\circle*{0.100000}}
\put(35.999996,27.104662){\circle*{0.100000}}
\put(37.499996,27.722240){\circle*{0.100000}}
\put(38.999996,28.335062){\circle*{0.100000}}
\put(40.499992,28.943659){\circle*{0.100000}}
\put(41.999992,29.548481){\circle*{0.100000}}
\put(43.499992,30.149920){\circle*{0.100000}}
\put(44.999989,30.748314){\circle*{0.100000}}
\put(46.499989,31.343958){\circle*{0.100000}}
\put(47.999985,31.937111){\circle*{0.100000}}
\put(49.499985,32.527996){\circle*{0.100000}}
\put(50.999985,33.116817){\circle*{0.100000}}
\put(52.499981,33.703747){\circle*{0.100000}}
\put(0.000000,0){\line(0,1){1}}
\put(15.000000,0){\line(0,1){1}}
\put(30.000000,0){\line(0,1){1}}
\put(45.000000,0){\line(0,1){1}}
\put(0.000000,0){\line(0,1){1}}
\put(0.000000,-3.5){\normalsize 0}
\put(15.000000,-3.5){\normalsize 1}
\put(30.000000,-3.5){\normalsize 2}
\put(45.000000,-3.5){\normalsize 3}
\put(0,10.000000){\line(1,0){1}}
\put(\xw,10.000000){\line(-1,0){1}}
\put(0,20.000000){\line(1,0){1}}
\put(\xw,20.000000){\line(-1,0){1}}
\put(0,30.000000){\line(1,0){1}}
\put(\xw,30.000000){\line(-1,0){1}}
\put(0,40.000000){\line(1,0){1}}
\put(\xw,40.000000){\line(-1,0){1}}
\put(0,10.000000){\line(1,0){1}}
\put(\xw,10.000000){\line(-1,0){1}}
\put(-1.900000,8.500000){\normalsize 0}
\put(-1.900000,18.500000){\normalsize 1}
\put(-1.900000,28.500000){\normalsize 2}
\put(-1.900000,38.500000){\normalsize 3}
{\linethickness{0.25mm}
\put(  0,  0){\line(1,0){\xw}}
\put(  0,\yw){\line(1,0){\xw}}
\put(\xw,  0){\line(0,1){\yw}}
\put(  0,  0){\line(0,1){\yw}} }
\end{picture} 
\vspace{-1cm}
\caption{Heavy quark potential at about $a^2\sigma \simeq 1.51$. 
The circles are measured values and dots are fitting values, 
for (a) Wilson, (b) Symanzik, (c) Iwasaki and (d) QCDTARO.}
\label{fig:potential}
\end{figure}
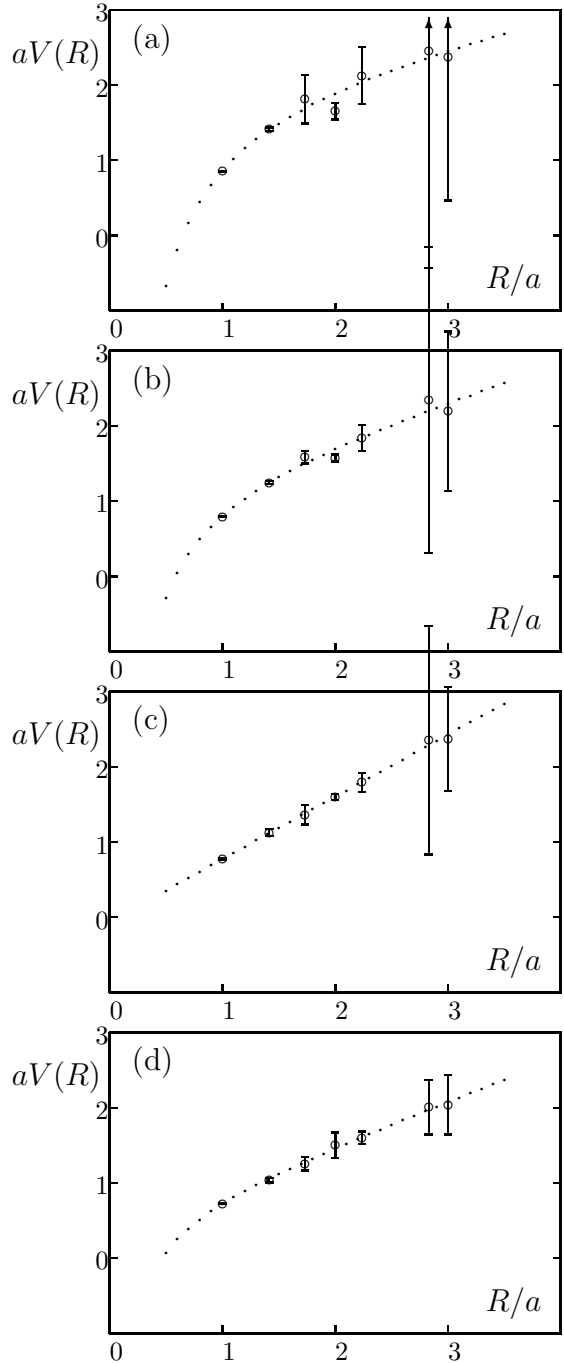

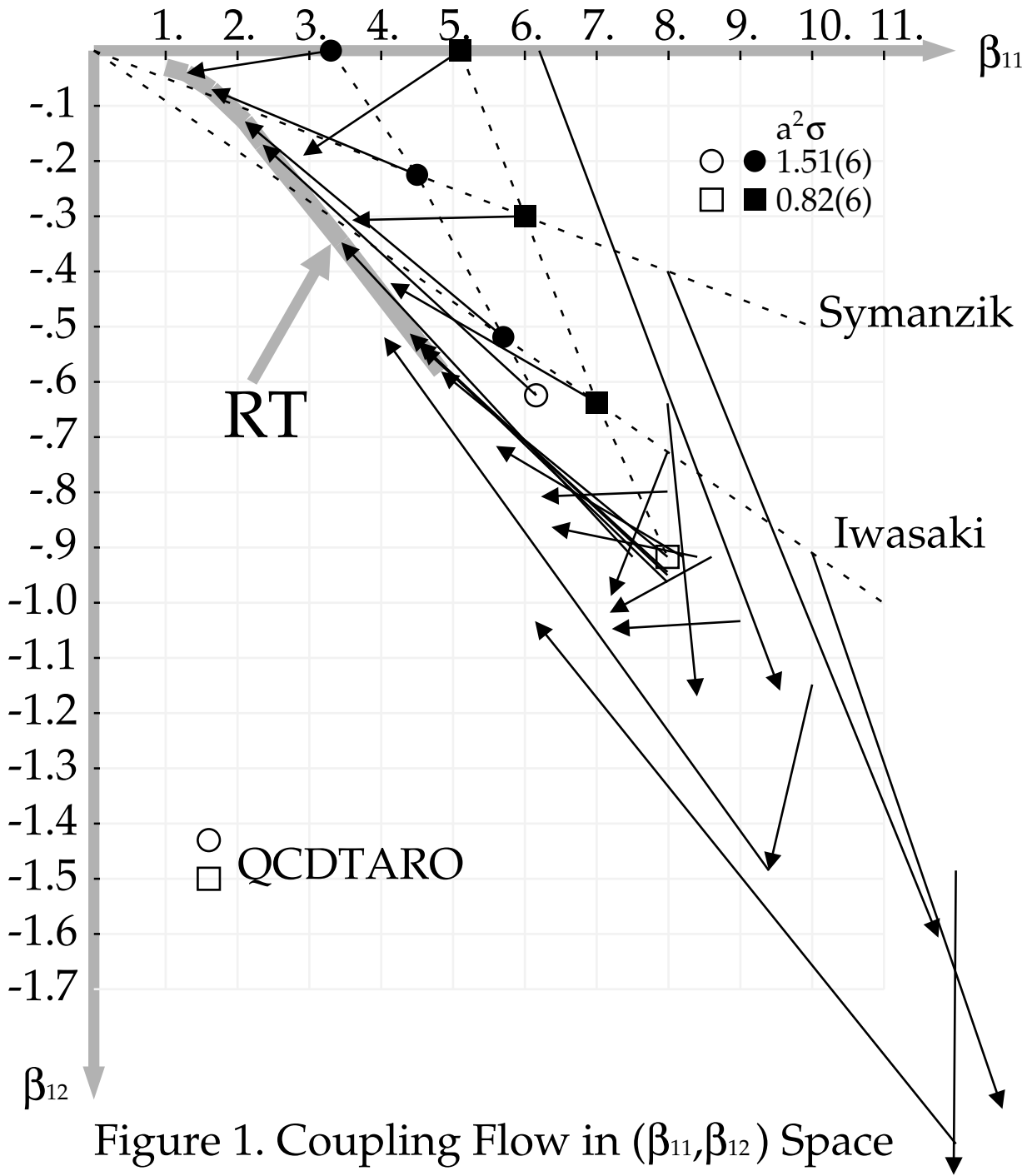
\end{document}